\documentclass[preprint,12pt]{elsarticle}

\usepackage{amssymb}
\usepackage{amsmath}

\journal{Nuclear Physics B}

\begin{document}

\begin{frontmatter}



\title{A Unified Variational Framework for Planar Elastica with General Distributed Loads}


\author{Yimu Mao}
\affiliation{organization={Stiftung Louisenlund},
            addressline={nimoseminov@gmail.com}, 
            city={Güby},
            postcode={24357}, 
            state={Schleswig-Holstein},
            country={Germany}}

\author{Christopher Tropp}

\affiliation{organization={Stiftung Louisenlund},
           addressline={christopher.tropp@louisenlund.de}, 
            city={Güby},
            postcode={24357}, 
            state={Schleswig-Holstein},
            country={Germany}}
\begin{abstract}
We present a simple variational framework for planar elastica that enables distributed energies—such as gravitational loading or magnetic body torques—to be incorporated in a modular and unified manner. The formulation is based on expressing all load-induced contributions directly at the level of the energy functional, which avoids the force-balance constructions used in classical treatments such as Wang (1986) and makes the inclusion of additional physical effects straightforward. The resulting planar energy functional yields compact governing equations in which the contributions of individual load types remain clearly separated. We demonstrate that the framework reproduces the classical heavy-elastica equations exactly and naturally accommodates magnetic-energy terms commonly used in hard-magnetic rod models. Although mathematically elementary, the formulation provides a clean and extensible structure for describing planar rod deformations under general distributed loads.
\end{abstract}


\end{frontmatter}

\section{Introduction}
\label{sec1}
Elastica is a fundamental model in mechanics for describing slender rods or beams, where the geometry is determined by the tangent angle or curvature distribution, and the mechanical behavior is typically characterized by bending energy and applied loads. Classical Euler elastica theory provides a systematic framework for studying deformations of rods in both plane and space. In many practical situations, rods experience distributed potential energies along their length—such as gravity, magnetic torques, or other body forces—leading to bending or complex planar deformations. Existing theories, however, are often tailored to specific types of loading and become cumbersome when multiple potentials are coupled, lacking a unified approach.
To address this challenge, we apply Fubini's theorem and thereby introduce a simple and general variational framework for planar elastica under arbitrary distributed potentials. The framework expresses all contributions at the level of the energy functional, avoiding the intricate derivations required in traditional force-balance methods. Any distributed potential—gravity, magnetic effects, or other body forces that may arise—can be incorporated in a unified manner, automatically yielding the corresponding Euler–Lagrange governing equations.
Importantly, when applied to planar rods subjected only to gravity, our framework exactly recovers the classical two-dimensional heavy-elastica results of Wang (1986), including cantilevered, vertical, and hanging rod configurations. Similarly, when applied to planar rods under only magnetic loading, the framework rigorously reproduces the planar deformation results reported by T.G. Sano, M. Pezzulla, P.M. Reis, et al. (2022). This demonstrates that our approach can naturally and consistently handle multiple physical potentials, providing a clear and general tool for analyzing and designing complex planar elastic-rod structures.

\section{Integral-Reduction Method for Nested Load Integrals}

We consider an inextensible planar rod parametrized by the arc-length $s\in[0,L]$. Many distributed loadings—such as magnetic body torques or gravity-induced forces—give rise to energy contributions that naturally involve nested integrals of the general form
\begin{equation}
\label{eq:nested_integral_general}
I := \int_{0}^{L} g_1(s_1) \Big( \int_{0}^{s_1} g_2(s_2)\, ds_2 \Big) ds_1,
\end{equation}
where $g_1$ and $g_2$ are continuous functions determined by the rod geometry and the applied fields.

The essential idea is to treat the nested integral \eqref{eq:nested_integral_general} as a structural object rather than focusing on elementary manipulation steps. This is Fubini's theorem. The integration domain
\[
\{0 \le s_2 \le s_1 \le L\} \subset \mathbb{R}^2
\]
is triangular, which naturally allows the introduction of a \emph{cumulative function} 
\begin{equation}
\label{eq:cumulative_function}
G(s) := \int_{s_2}^L g_1(\xi)\, d\xi,
\end{equation}
and leads to a compact, single-integral representation
\begin{equation}
\label{eq:single_integral_general}
I = \int_0^L G(s)\, g_2(s)\, ds.
\end{equation}

This formulation isolates the structural coupling encoded by the nested integral and provides a systematic framework to simplify energy functionals for planar rods subject to distributed loads. The same viewpoint extends naturally to multi-layer nested integrals. For instance, an $n$-fold nested integral of the form
\begin{equation}
\int_0^L g_1(s_1) \int_0^{s_1} g_2(s_2) \cdots \int_0^{s_{n-1}} g_n(s_n)\, ds_n \cdots ds_2 ds_1
\end{equation}
can be recursively reduced by defining a hierarchy of cumulative functions $G_k(s)$, yielding a compact or single-integral expression. In this work, we focus on the two-layer case $n=2$, while noting that the framework applies verbatim to higher-order nested integrals encountered in rod mechanics and related variational problems.

\section{Application to Hard Magnetic Rods under Planar Deformations}
\label{sec:hardmagnetic}

We now apply the mathematical framework to analyze the planar deformation of hard magnetic rods under external magnetic fields. Thereby, we follow the work of ~\cite{Sano2022}. Consider an inextensible rod of length $L$ with circular cross-section of radius $R$, uniformly magnetized along its centerline with residual flux density $B^r$. $\boldsymbol{F}$ describes the deformation gradient. The total energy functional combines elastic bending and magnetic contributions:

\begin{equation}
E_{\text{total}} = \int_0^L \frac{EI}{2}(\theta')^2  ds - \frac{A}{\mu_0} \int_0^L (\mathbf{F}\mathbf{B}^r) \cdot \mathbf{B}^a  ds,
\end{equation}

where $A = \pi R^2$ is the cross-sectional area. Assuming the magnetic moment aligns with the tangent direction $\mathbf{t}(s) = (\cos\theta(s), \sin\theta(s))$ and the rod is inextensible (so the deformation gradient factor $F = 1$), the magnetic energy simplifies to:

\begin{equation}
E_{\text{mag}} = -\frac{A B^r}{\mu_0} \int_0^L \left[ B_x^a(s) \cos\theta(s) + B_y^a(s) \sin\theta(s) \right] ds.
\end{equation}

To handle the nested integrals arising from the variation, we use the integral representations of the trigonometric functions. Specifically, the tangent angle $\theta(s)$ can be expressed as:

\begin{equation}
\theta(s) = \theta(0) + \int_0^s \theta'(\tau)  d\tau,
\end{equation}

which allows us to expand $\sin\theta(s)$ and $\cos\theta(s)$ as:

\begin{equation}
\sin\theta(s) = \sin\theta(0) + \int_0^s \theta'(\tau) \cos\theta(\tau)  d\tau, \quad \cos\theta(s) = \cos\theta(0) - \int_0^s \theta'(\tau) \sin\theta(\tau)  d\tau.
\end{equation}

Substituting these into the magnetic energy and applying Fubini's theorem to exchange the order of integration, we define cumulative field functions:

\begin{equation}
K_x^a(\tau) = \int_\tau^L B_x^a(s)  ds, \quad K_y^a(\tau) = \int_\tau^L B_y^a(s)  ds.
\end{equation}

After simplification, the magnetic energy reduces to a single integral:

\begin{equation}
E_{\text{mag}} = -\frac{A B^r}{\mu_0} \int_0^L \theta'(\tau) \left[ -\sin\theta(\tau) K_x^a(\tau) + \cos\theta(\tau) K_y^a(\tau) \right] d\tau + \text{constant}.
\end{equation}

The corresponding Euler--Lagrange equation yields the general governing equation:

\begin{equation}
EI \frac{d^2 \theta}{ds^2} - \frac{A B^r}{\mu_0} \left[ -\sin\theta(s) K_x^a(s) + \cos\theta(s) K_y^a(s) \right] = 0.
\label{eq:hardmagnetic}
\end{equation}

\subsection{Reduction to the Constant Gradient Field Case}

We now demonstrate how equation \eqref{eq:hardmagnetic} reduces to the classical form for a constant gradient magnetic field $\mathbf{B}^a = b y \hat{\mathbf{e}}_y$. For this field, we have:

\begin{equation}
B_x^a = 0, \quad B_y^a = b y(s), \quad K_x^a(s) = 0, \quad K_y^a(s) = b \int_s^L y(s')  ds'.
\end{equation}

Substituting into equation \eqref{eq:hardmagnetic} gives:

\begin{equation}
EI \frac{d^2 \theta}{ds^2} - \frac{A B^r b}{\mu_0} \cos\theta(s) \int_s^L y(s')  ds' = 0.
\end{equation}

Applying Fubini's theorem and relabeling dummy variables as needed, the integral can be simplified, yielding the final planar form:

\begin{equation}
EI \frac{d^2 \theta}{ds^2} - \frac{A B^r b}{\mu_0} \left[ y(s) \sin\theta(s) - \cos\theta(s) \int_s^L \cos\theta(s')  ds' \right] = 0.
\label{eq:finalplanar}
\end{equation}

This final expression is consistent with the planar version of the governing equation reported by Sano et al. \cite{Sano2022}, thereby confirming the validity of the simplification framework proposed in this section.

\subsection{Extension to the \(X\!-\!Z\) Plane with Gravitational Loading}

To extend the preceding \(X\!-\!Y\) planar formulation to the \(X\!-\!Z\) plane, 
we simply replace all occurrences of the transverse coordinate \(y(s)\) with 
\(z(s)\). The tangent direction becomes 
\(\mathbf{t}(s) = (\cos\theta(s),\, \sin\theta(s))\) in the \(X\!-\!Z\) plane, 
and the kinematic relations read
\begin{equation}
x'(s)=\cos\theta(s),\qquad 
z'(s)=\sin\theta(s).
\end{equation}

We now incorporate gravitational potential energy. For a rod of mass density 
\(\rho\) and cross-sectional area \(A\), gravity acts in the negative \(Z\)-direction, 
and the gravitational potential energy is
\begin{equation}
E_{\mathrm{grav}} = \rho A g \int_{0}^{L} z(s)\, ds,
\end{equation}
where \(g\) is the gravitational acceleration. 
The corresponding variational contribution involves nested integrals through
\begin{equation}
z(s)=z(0)+\int_{0}^{s}\sin\theta(\sigma)\, d\sigma.
\end{equation}
Applying the integral-reduction method introduced in Section~\ref{sec1}, 

we express the gravitational term in the single-integral form
\begin{equation}
E_{\mathrm{grav}}
= \rho A g \int_{0}^{L} 
\left( \int_{s}^{L} ds' \right)\sin\theta(s)\, ds
= \rho A g \int_{0}^{L} (L-s)\,\sin\theta(s)\, ds.
\end{equation}

Combining bending, magnetic, and gravitational contributions yields the 
energy functional
\begin{equation}
E_{\mathrm{tot}}
= \int_{0}^{L} \frac{EI}{2}(\theta')^{2} ds 
- \frac{A B^{r}}{\mu_{0}}
  \int_{0}^{L}\!\big[
    B_x^{a}(s)\cos\theta(s)
    + B_z^{a}(s)\sin\theta(s)
  \big] ds
+ \rho A g \int_{0}^{L} (L-s)\sin\theta(s)\, ds.
\end{equation}

Introducing the cumulative-field functions
\begin{equation}
K_x^{a}(s)=\int_{s}^{L} B_x^{a}(s')\, ds',\qquad
K_z^{a}(s)=\int_{s}^{L} B_z^{a}(s')\, ds',
\end{equation}
and applying the same nested-integral reduction as before, the Euler–Lagrange 
equation becomes
\begin{equation}
EI\,\theta''(s)
-\frac{A B^{r}}{\mu_{0}}
  \big[
    -\sin\theta(s)\, K_x^{a}(s)
    +\cos\theta(s)\, K_z^{a}(s)
  \big]
+\rho A g (L-s)\cos\theta(s)
=0.
\end{equation}

This equation constitutes the \(X\!-\!Z\)-planar analog to the earlier 
\(X\!-\!Y\) formulation, now with gravitational loading included. 
It preserves the compact single-integral structure enabled by the 
integral-conversion method.

\section{Application to the 2D Heavy Elastica}

To extend the planar analysis developed in the \(X\!-\!Y\) setting to the \(X\!-\!Z\) plane, it suffices to replace the transverse coordinate \(Y\) by \(Z\) in all expressions for the geometry, kinematics, and elastic energy. The only additional contribution is the gravitational potential, which for an inextensible rod takes the classical form  
\begin{equation}
E_g=\int_0^L \lambda g\, Z(s)\, ds,
\end{equation}
whose first variation simply adds a uniform body force \(\lambda g\) along the negative \(Z\)-direction. Consequently, the Euler–Lagrange equation obtained in our unified integral-reduction framework becomes  
\begin{equation}
\label{eq:heavy_EL}
EI\,\theta''(s)=\lambda g (L-s)\cos\theta(s)-F_1 \sin\theta(s),
\end{equation}
where \(\theta(s)\) is the tangential angle of the centerline and \(F_1\) is the internal force component determined by the boundary conditions.

Equation \eqref{eq:heavy_EL} directly recovers the classical heavy elastica models.  
Indeed, under the standard choice of the force resultants for a cantilevered or hanging rod, \eqref{eq:heavy_EL} can be rewritten into the form derived by Wang~\cite{Wang}:
\begin{equation}
\theta''(s)=H\sin\theta(s)-(V+Bs)\cos\theta(s),
\end{equation}
upon identifying
\begin{equation}
H=\frac{-F_1}{EI},\qquad 
V=\frac{-\lambda g L}{EI},\qquad 
B=\frac{\lambda g}{EI}.
\end{equation}
This agreement shows that the classical scenarios analyzed by Wang ~\cite{Wang}—including the heavy cantilever, the standing heavy column, and the long hanging column—are all naturally encompassed within our variational formulation.  
Therefore, the present framework not only accommodates external magnetic loads but also reproduces the full spectrum of two-dimensional heavy elastica configurations, confirming the correctness and robustness of our method.

\begin{figure}[h]
    \centering
    \begin{minipage}{0.5\linewidth}
        \centering
        \includegraphics[width=\linewidth]{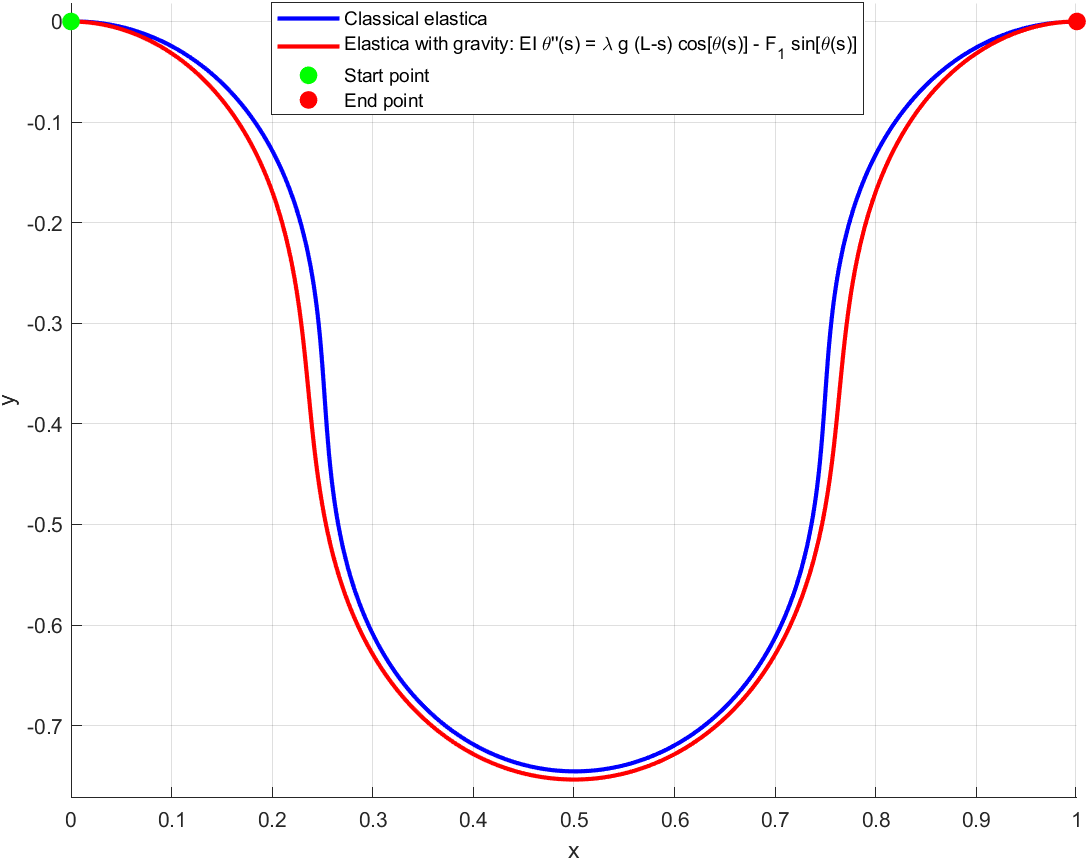}
        \caption{Comparison of the corrected equation with the classical Euler elastica}
        \label{fig:elastica}
    \end{minipage}

\end{figure}

\section{Conclusion}

This work describes an general integral simplification method for planar elastic rods, which reduces nested integrals arising from distributed loads to compact single-integral expressions, thereby simplifying the analysis and computation of the energy functional. The method is validated in the contexts of planar hard-magnetic rods and two-dimensional heavy elastic bodies, where it successfully reproduces classical results and enables the convenient incorporation of magnetic and gravitational effects. It should be emphasized that the present approach applies only to planar deformations and cannot be directly extended to fully three-dimensional configurations. In addition, the accumulated field functions such as $B_x^a$ are assumed to be known a priori, so the resulting governing differential equations contain no nested integrals.

\end{document}